\documentclass[%
reprint,
groupedaddress,
showpacs,preprintnumbers,
 amsmath,amssymb,
 aps,
 prl,
]{revtex4-1}

\usepackage{bbm}
\usepackage{graphicx}
\usepackage{dcolumn}
\usepackage{bm}
\usepackage{amssymb}
\usepackage{pifont}
\usepackage{hyperref}

\begin{document}

\preprint{APS/123-QED}

\title{Motional-Mode Analysis of Trapped Ions}

\author{Henning  Kalis}
\email{henning.kalis@physik.uni-freiburg.de}
\author{Frederick Hakelberg}%
\author{Matthias Wittemer}%
\author{Manuel Mielenz}%
\author{Ulrich Warring}%
\author{Tobias Schaetz}%
\email{tobias.schaetz@physik.uni-freiburg.de}
\affiliation{ Albert-Ludwigs-Universit\"at Freiburg, Physikalisches Institut, Hermann-Herder-Stra\ss e 3, 79104 Freiburg, Germany}
\date{\today}

\begin{abstract}
We present two methods for characterization of motional-mode configurations that are generally applicable to the weak and strong-binding limit of single or multiple trapped atomic ions. 
Our methods are essential to realize control of the individual as well as the common motional degrees of freedom.
In particular, when implementing scalable radio-frequency trap architectures with decreasing ion-electrode distances, local curvatures of electric potentials need to be measured and adjusted precisely, e.g., to tune phonon tunneling and control effective spin-spin interaction.
We demonstrate both methods using single $^{25}$Mg$^+$ ions that are individually confined $40\,\mu$m above a surface-electrode trap array and prepared close to the ground state of motion in three dimensions. 

\begin{description}
\item[PACS: 37.10.Ty, 45.10.Na]
\end{description}
\end{abstract}

\maketitle


Quantum simulators\,\cite{feynman_simulating_1982,*buluta_quantum_2009,*schaetz_focus_2013,*georgescu_quantum_2014,cirac_goals_2012} and hybrid quantum systems\,\cite{xiang_hybrid_2013} may present ideal platforms to experimentally explore the emergence and dynamics of complex quantum phenomena, such as many body physics in strongly-correlated systems\,\cite{eckert_quantum_2008,cirac_goals_2012} or geometrical frustration\,\cite{moessner_two-dimensional_2000}.
Promising candidates for realizations of scalable quantum simulators, implementing coherent control between constituents, are based on atoms\,\cite{lewenstein_ultracold_2007}, photons\,\cite{hartmann_strongly_2006}, electrons\,\cite{hanson_coherent_2008}, or atomic ions\,\cite{porras_quantum_2006,schneider_experimental_2012}.
Hybrid quantum systems combine two or more different physical constituents, e.g., spins\,\cite{nowack_coherent_2007}, atoms\,\cite{saffman_quantum_2010,*deng_storage_2010}, or solid state devices\,\cite{togan_quantum_2010} and pursue studying mutual interactions. 
Both platforms benefit from advances in micro fabrication techniques\,\cite{daniilidis_fabrication_2011,*sterling_fabrication_2014}, yielding increasing interaction strengths by decreasing system length scales.
Correspondingly, higher-order terms need to be considered, in order to enable precise control of interaction potentials.
For example, in microfabricated surface-electrode ion trap arrays\,\cite{seidelin_microfabricated_2006,schmied_optimal_2009,mielenz_freely_2015} local potentials, dominated by applied electric trapping potentials, define motional modes.
For envisioned quantum simulations, motional degrees of freedom can be exploited either within individual sites or between different sites.
This, in turn, requires adjustment of motional-mode configurations, i.e., individual orientation of the normal-mode vectors and related motional frequencies, to enable individual, tunable inter-site interactions\,\cite{mielenz_freely_2015}.
In this letter, we introduce and experimentally demonstrate two distinct methods for the analysis of motional-mode configurations that are generally applicable to the weak and strong-binding limit. 
For the latter, we cool single ions close to the ground state of motion in three dimensions.

\begin{figure}
\includegraphics[width=\columnwidth]{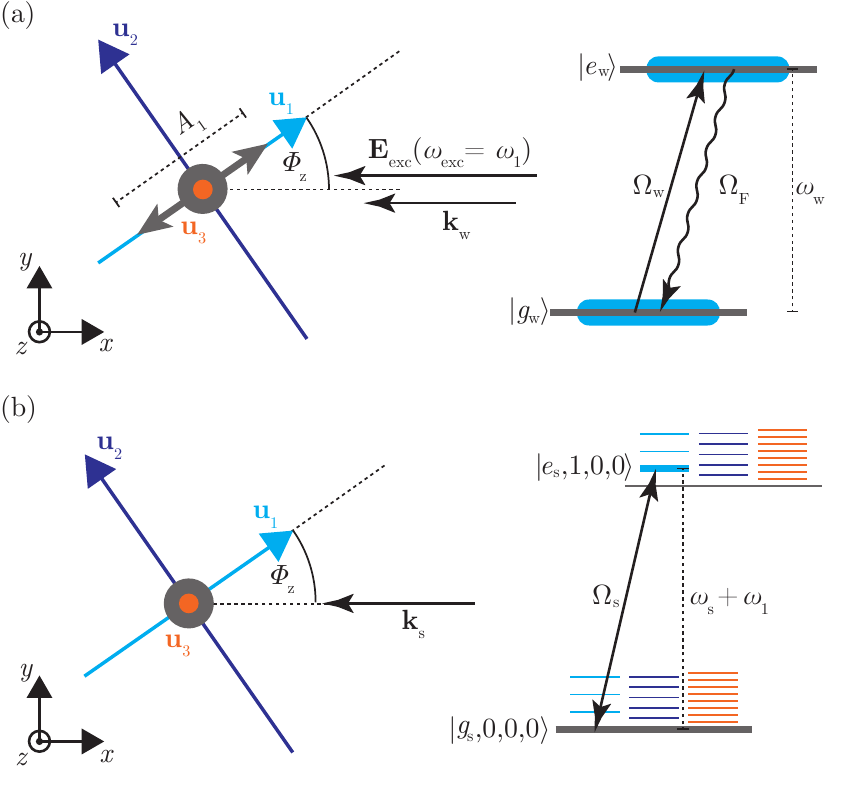}%
\caption{
(color online)
Simplified illustration of both mode-analysis methods applicable to the weak and strong-binding limit. In our simplified example, the motional modes are rotated relative to the coordinate system by $\phi_{\rm{z}}$. 
(a) In the weak-binding limit, the mode $\mathbf{u}_{1}$ is resonantly excited by an electric field $\mathbf{E}_{\text{exc}}$ oscillating at $\omega_{\rm{exc}} = \omega_1$ to motional amplitude $A_1\propto \langle \mathbf{u}_1 , \mathbf{E}_{\text{exc}}\rangle$ with $\langle n_1 \rangle \gg 0$. 
Subsequently, a laser with wave vector $\mathbf{k}_{\rm{w}}$ along $x$ resonantly drives the transition $\left|g_{\rm{w}} \right \rangle \leftrightarrow \left|e_{\rm{w}} \right \rangle$. Blue bars illustrate the natural line width $\Gamma_{\rm{w}}$. 
In turn, the mode orientation is derived considering the Doppler effect and the related decrease of the fluorescence rate $\Omega_{\text{F}}$ as a function of $ A_1 \langle\mathbf{u}_{1},\mathbf{k}_{\rm{w}}\rangle$.
(b) In the strong-binding limit, a laser is tuned to $\omega_{\rm{s}}+\omega_1$ with wave vector $\mathbf{k}_{\rm{s}}$ along $x$, coherently coupling the electronic (gray bars) and motional states (colored bars), e.g., $|g_{\rm{s}} ,0,0,0\rangle$ to $|e_{\rm{s}} ,1,0,0\rangle$.
Here, the coupling rate $\Omega_{\rm{s}} \propto \langle e_{\rm{s}} , 1, 0, 0|e^{i\langle\mathbf{k}_{\rm{s}},\mathbf{u}_1\rangle}|g_{\rm{s}} ,0,0,0\rangle$ encodes the mode configuration. 
\label{fig1}}
\end{figure}

To introduce our system, we consider a single ion with charge $Q$ and mass $m$, harmonically bound in three dimensions with motional frequencies $\omega_{i}$ along normalized mode vectors $\mathbf{u}_{i}$, where mode $i \in \{1,2,3\}$.
Initially, modes are arbitrarily rotated with respect to the axes of the laboratory frame, labeled $x$, $y$, and $z$, and we parametrize mode orientations by three subsequent rotations around these spatially-fixed axes by the angles $\phi_x$, $\phi_y$, and $\phi_z$, with the corresponding composite rotation matrix $\mathcal{R}(\phi_x,\phi_y,\phi_z)$.
Further, motional states are described by Fock state distributions $\mathcal{P}_i(n_i)$, where $n_i$ are the corresponding phonon numbers of mode $i$.
We consider one electronic ground state $\left|g_\lambda\right\rangle$ and one excited state $\left|e_\lambda\right \rangle$ with energy difference $\hbar \omega_{\lambda} \gg \hbar \omega_i$ and a natural line width $\Gamma_{\lambda}$, with $\lambda \in \{ \rm{w},\rm{s}\}$.
For $\Gamma_{\rm{w}} \gg \omega_i$, the so called weak-binding limit, our first method is suited best to determine $\boldsymbol{\omega}/(2\pi) \equiv \{\omega_1, \omega_2, \omega_3\}/(2\pi) $ and $\{\phi_x, \phi_y, \phi_z\}$. In Figure \ref{fig1}(a) we show a simplified illustration for $\{\phi_x, \phi_y, \phi_z\} \approx \{0, 0, 35\}^{\circ}$.
%
%
The method is based on the excitation of the modes by an electric field $\mathbf{E}_{\text{exc}}$ oscillating at $\omega_{\text{exc}} \approx \omega_i$ for  duration $t_{\text{exc}}$.
Motional amplitudes $A_i(t_{\text{exc}}) \propto \langle \mathbf{u}_i , \mathbf{E}_{\text{exc}}\rangle$ along $\mathbf{u}_i$ can be evaluated classically, when excited to a coherent state with $\langle n_ i \rangle \equiv \langle\mathcal{P}_i(n_i)\rangle \gg 0$.
Subsequently, the fluorescence rate $\Omega_{\text{F}}$ induced by a laser tuned to $\omega_{\rm{w}}$ with wave vector $\mathbf{k}_{\rm{w}}$ and rate $\Omega_{\rm{w}}< \Omega_{\rm{F}}$ [see Fig.\,\ref{fig1}(a)], is modulated as a function of $A_i \langle\mathbf{u}_{i},\mathbf{k}_{\rm{w}}\rangle$ due to the Doppler effect\,\cite{jefferts_coaxial-resonator-driven_1995,*gudjons_influence_1997}.
In the following, we refer to a normalized fluorescence $\mathcal{F} \equiv \Omega_{\text{F}}/\Omega'_{\text{F}}$, where $\Omega'_{\text{F}}$ represents the unmodulated rate. 
In contrast, in the strong-binding limit, $\Gamma_{\rm{s}} \ll \omega_i$, we present our second method to determine mode configurations; we sketch the corresponding example in Fig.\,\ref{fig1}(b).
Here, motional excitation of single quanta is performed via a laser, exemplarily tuned to $\omega_{\rm{s}} + \omega_{1}$, coherently coupling the electronic and motional states at a rate $\Omega_{\rm{s}} \propto \langle e_{\rm{s}}, 1, 0, 0|e^{i\langle\mathbf{k}_{\rm{s}},\mathbf{u}_1\rangle}|g_{\rm{s}},0,0,0\rangle$\,\cite{leibfried_quantum_2003}.
We determine the angle between $\mathbf{k}_{\rm{s}}$ and $\mathbf{u}_i$ by evaluating $\Omega_{\rm{s}}$ for excitation at $\omega_{\rm{s}} + \omega_i$, which is efficiently performed when $\eta_i \sqrt{2\langle n_i\rangle+1} \ll 1$, with the Lamb-Dicke Parameter $\eta_i = \left|\langle\mathbf{k}_{\rm{s}},\mathbf{u}_i\rangle\right| \sqrt{\hbar/(2m\omega_i)}$.
Application of either method yields unique results for arbitrary mode configurations, when sequentially probed from multiple directions.

\begin{figure}
\includegraphics[width=\columnwidth]{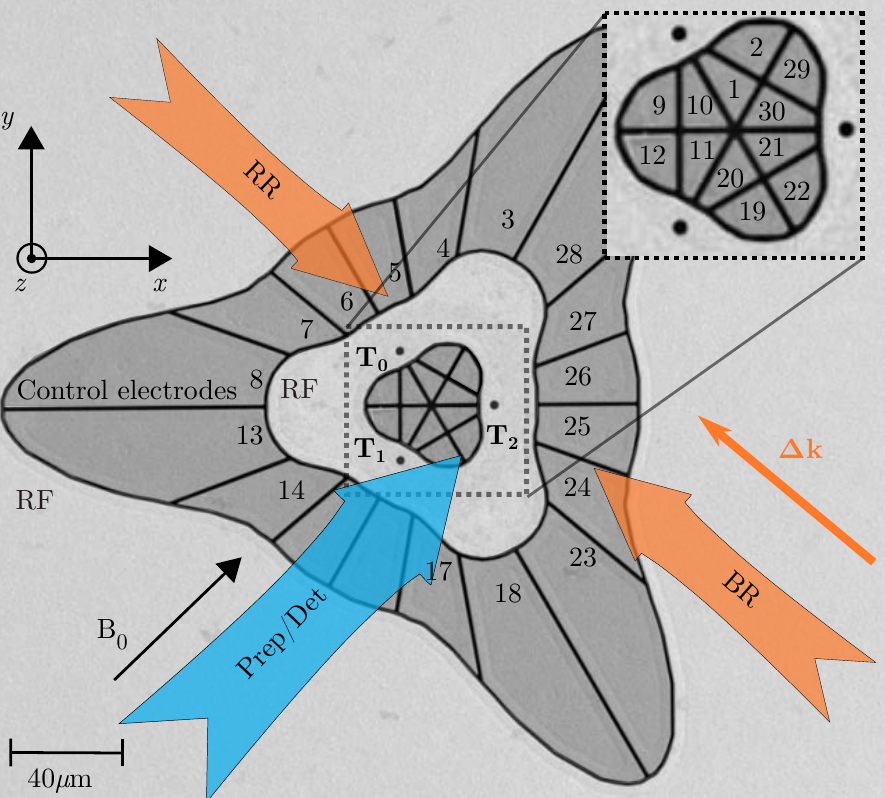}%
\caption{(color online)
Illustration of the experimental setup, by a false colored scanning electron microscope image, showing our surface-electrode trap array.
The array consists of two radio-frequency electrodes (light grey) and 30 control electrodes (dark grey), labeled $1 \dots 30$.
Trapping sites are denoted $\mathbf{T}_j$ with $j \in \{0,1,2\}$ and are separated by $\approx 40\,\mu$m; 
here, all experiments are carried out with single $^{25}$Mg$^{+}$ ions trapped near $\mathbf{T}_2$. 
Motional excitation fields $\mathbf{E}_{\rm{exc}} = U_{\rm{exc}}\mathbf{E}_l$ oscillating at $\omega_{\rm{exc}}$ are applied via the $l$-th control electrode.
A set of laser beams, propagating along $\mathbf{k}_{\rm{w}}$ and parallel to the magnetic field $|\mathbf{B}_0|\approx 4.65\,$mT, is used for preparation and detection of the electronic degrees of freedom (blue arrow).
Two more beams, labeled BR/RR (red arrows), with an effective wave vector $\Delta \mathbf{k} \equiv \mathbf{k}_{\rm{s}}$, are implemented to coherently couple the ion's electronic and motional states via two-photon stimulated Raman transitions.
\label{fig2}}
\end{figure}

We experimentally demonstrate both methods with single $^{25}\text{Mg}^{+}$ ions harmonically confined in a surface-electrode trap array fabricated by Sandia National Laboratories\,\cite{tabakov_assembling_2015,mielenz_freely_2015}.
The array inherits a triangular arrangement of individual traps with an inter-site distance of $d \approx 40$\,$\mu$m, see Fig. \ref{fig2}.
%
%
Three-dimensional confinement at three distinct trapping sites is realized by a radio-frequency (RF) potential $\phi_{\text{RF}}(\mathbf{r})$, oscillating at  $\Omega_{\text{RF}}/(2\pi) = 88.1\,$MHz with a peak voltage $U_{\text{RF}} \approx 50$\,V applied in phase to two RF electrodes.
We refer to the trapping sites as $\mathbf{T}_j$ with $j \in \{0,1,2\}$ and, e.g., $\mathbf{T}_2$ located at $\{24, 0,  36\}\,\mu$m.
Typical motional frequencies are $\boldsymbol{\omega}/(2\pi) \approx \{3.6, 4.8, 5.9\}$\,MHz.
Our array provides 30 control electrodes that we offset to a constant voltage $\in [-10, 10]$\,V to control the motional degrees of freedom\,\cite{mielenz_freely_2015}. 
The $l$-th electrode biased at 1\,V generates a potential $\phi_{l}(\mathbf{r})$ with $l \in \{1,\dots,30\}$, which is evaluated, e.g., at $\mathbf{r} = \mathbf{T}_2$, using the gapless-plane approximation\,\cite{Hucul:2008:TAI:2016976.2016977,*wesenberg_electrostatics_2008,*schmied_electrostatics_2010}.
For the $l$-th electrode of choice we can create oscillating $\phi_{l}(\mathbf{r})$ by applying a signal from a direct digital synthesizer (DDS) at $\omega_{\text{exc}}/(2\pi) \approx$ 1--10\,MHz and peak voltages $U_{\text{exc}} \approx$ 0.1--1\,mV.

For preparation, manipulation and detection of the electronic and motional states we employ laser beams with wavelengths close to $280\,$nm\,\cite{friedenauer_high_2006}.
All beams pro\-pa\-gate parallel to the surface (${xy}$-plane).
Four overlapping $\sigma^+$-polarized beams, propagating parallel to a static magnetic field $|\mathbf{B}_0| \approx 4.65\,$mT, are used for Doppler cooling and electronic-state preparation.
One beam (BDD) is detuned by $\Delta_{\text{BDD}}/(2\pi)  \approx -8\Gamma_{\text{w}}/(2\pi)$ from the cycling transition $\left|g\right\rangle \equiv \left|g_{\{\rm{w,s}\}}\right\rangle = \left|S_{1/2}, F = 3, m_F = +3 \right \rangle \leftrightarrow \left| e_{\text{w}}\right\rangle \equiv \left|P_{3/2}, F = 4, m_F = +4 \right \rangle$ with a natural line width of $\Gamma_{\text{w}}/(2\pi) \approx 42\,\rm{MHz}$.
The other beam (BD) with wave vector $\mathbf{k}_{\text{BD}}\equiv \mathbf{k}_{\rm{w}}$ is detuned by $\Delta_{\text{BD}}/(2\pi)  \approx -\Gamma_{\text{w}}/(4\pi)$ with intensity $I_{\rm{BD}} \approx I_{\text{sat}}/2$, where $ I_{\text{sat}} \approx 2500\,$W/m$^2$ denotes the saturation intensity; for state-dependent fluorescence detection\,\cite{leibfried_quantum_2003}, we switch $\Delta_{\text{BD}}/(2\pi) \approx -5\,$MHz.
Further, two optical-pumping beams are used to prepare $\left|g\right\rangle$.
We implement two more beams (BR/RR) with an effective wave vector $\Delta \mathbf{k}\equiv \mathbf{k}_{\rm{s}}$ along $\{-1/\sqrt{2}, 1/\sqrt{2},0\}$ to drive two-photon stimulated Raman transitions\,\cite{leibfried_quantum_2003}, with $\Gamma_{\rm{s}} \ll \Omega_{\rm{s}} \ll \omega_{\rm{s}}$, between $\left| g\right\rangle$ and $\left| e_{\rm{s}}\right\rangle \equiv \left|S_{1/2}, F = 2, m_F = +2 \right \rangle$, separated by $\omega_{\rm{s}}/(2\pi) \approx 1681.5\,$MHz.
The beams are polarized $1/\sqrt{2}(\sigma^+ + \sigma^-)$ and $\pi$, respectively, coupling $\left| g\right\rangle$ and $\left| e_{\rm{s}}\right\rangle$ via a common virtual level, detuned by $\Delta_{\text{R}}/(2\pi)  \approx 33\,$GHz from the $P_{3/2}$ manifold.
The relative detuning of the beams can be varied between $\omega_{\rm{R}}/(2\pi)\approx \omega_{\rm{s}}/(2\pi) \pm 40\,$MHz. 

\begin{figure}
\includegraphics[width=\columnwidth]{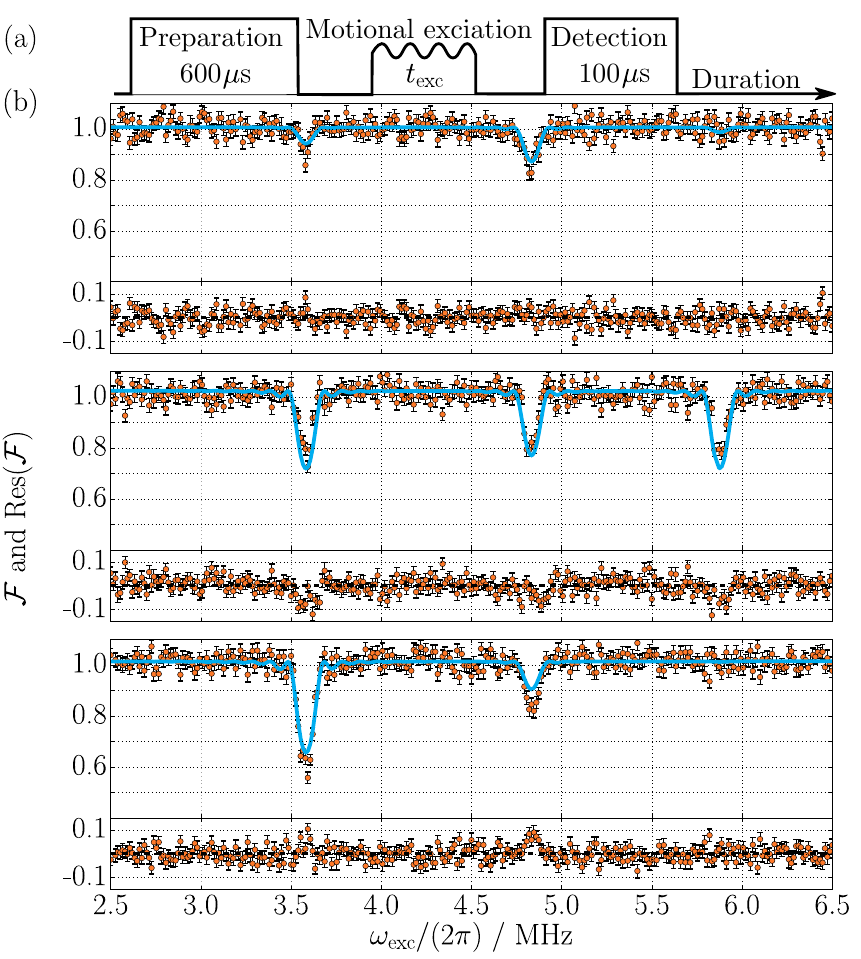}%
\caption{(Color online) Determination of mode configurations, i.e, motional frequencies and mode orientations, in the weak-binding limit. 
(a) Experimental sequences performed with a single ion trapped near $\mathbf{T}_2$, beginning with Doppler cooling and preparation of $|g\rangle$, followed by a motional excitation pulse of duration $t_{\text{exc}} = 10\,\mu$s, and finalized by the measurement of $\Omega_{\text{F}}$, revealing the normalized fluorescence $\mathcal{F}$. 
(b) Excitation pulses are applied using $\mathbf{E}_{\rm{exc}} \propto \mathbf{E}_l$, representatively shown for $l \in \{22,26,28\}$ (from top to bottom), with corresponding $\mathcal{F}$ as a function of the excitation frequency $\omega_{\text{exc}}$.
Data points are taken in random order and averaged over 200 repetitions of sequences with fixed parameter settings, where error bars denote the statistic uncertainties.
A combined model fit (solid lines) considering ten different electrodes $l \in \{21\dots 30\}$ yields mode configurations: $\{\phi_x, \phi_y, \phi_z\} = \{-6(1)_{\text{stat}}(3)_{\text{sys}}, -38(1)_{\text{stat}}(4)_{\text{sys}}, -1(1)_{\text{stat}}(1)_{\text{sys}}\}^{\circ}$ and $\boldsymbol{\omega}/(2\pi) = \{3.584(2)_{\rm{stat}}, 4.833(3)_{\rm{stat}}, 5.878(4)_{\rm{stat}}\}$\,MHz. Residuals are shown below each spectrum.
}
\label{fig3}
\end{figure}

To demonstrate the first method we apply the experimental sequence, illustrated in Fig.\,\ref{fig3}(a), to a single ion trapped near $\mathbf{T}_2$: 
After Doppler cooling and preparation of $|g\rangle$, the DDS signal at $\omega_{\rm{exc}}$ is capacitively coupled onto one pre-selected control electrode for duration $t_{\text{exc}} = 10\,\mu$s. 
Finally, we detect the fluorescence on the transition $|g\rangle \leftrightarrow |e_{\rm{w}}\rangle$ for a duration of $100\,\mu$s. 
In subsequent measurements, we choose the ten electrodes $l \in \{21\dots 30\}$ with $\mathbf{E}_{\rm{exc}} = U_{\rm{exc}}\,\mathbf{E}_{l} \equiv - U_{\rm{exc}}\nabla\,\phi_{l}(\mathbf{r})|_{\mathbf{r}=\mathbf{T}_2}$, record $\mathcal{F}$ as a function of $\omega_{\text{exc}}$, subtract the independently determined stray-light contributions from the total photon counts, and average each data point over 200 repetitions.
In Figure\,\ref{fig3}(b), we show, as an example, recorded $\mathcal{F}$ for selected electrodes $l \in \{22,26,28\}$.
%
%

For the analysis, we consider that the motional modes $\mathbf{u}_i$ remain uncoupled and introduce a model for the modulation of $\mathcal{F}$, based on the classical driven harmonic oscillator\,\cite{kalis_motional_supplement_2016}. Here, final motional amplitudes are given by:
\begin{equation}
A_i(t_{\text{exc}}) =\frac{Q}{m}U_{\text{exc}}\left |\left\langle\mathbf{u}_i,\mathbf{E}_{l}\right\rangle \right| \frac{2}{\omega_{\text{exc}}^2 - \omega_{i}^2} \sin\left[\frac{t_{\text{exc}}}{2}\left(\omega_{\text{exc}} - \omega_{i}\right)\right],
\end{equation}
where we assume that the ion is initially at the center of the harmonic potential and at rest. 
Further, motional excitations result in\,\cite{devoe_role_1989,*berkeland_minimization_1998}
\begin{equation}
\mathcal{F} = \left(\frac{\Gamma_{\text{w}}}{2}\right)^2\,\displaystyle\prod_{i=1}^{3} \sum_{v = -\infty}^{\infty}\frac{J_v(\beta_{i})^2}{(\Delta_{\text{BD}} + v\,\omega_{i})^2 + (\Gamma_{\text{w}}/2)^2} \, \text{,}
\label{sec:WBL:eq1}
\end{equation}
%
%
where we use the $v$-th Bessel function $J_v$ and the modulation index $\beta_{i} =\left|\left\langle\mathbf{u}_i,\mathbf{k}_{\text{BD}}\right\rangle\right| A_i(t_{\text{exc}})$. 
In a combined fit of this model to all ten spectra, with seven free parameters, we obtain $\{\phi_x, \phi_y, \phi_z\}\,=\,\{-6(1)_{\text{stat}}(3)_{\text{sys}}, -38(1)_{\text{stat}}(4)_{\text{sys}}, -1(1)_{\text{stat}}(1)_{\text{sys}}\}^{\circ}$, $U_{\text{exc}} = 660(10)_{\rm{stat}}\,\mu$V, and $\boldsymbol{\omega}/(2\pi) = \{3.584(2)_{\rm{stat}}, 4.833(3)_{\rm{stat}}, 5.878(4)_{\rm{stat}}\}$\,MHz, where we truncate the sum in Eq.\,\eqref{sec:WBL:eq1} at $|v| = 15$.
The mode configuration is uniquely determined by the fit and from the parameters we calculate the (quasi-static) curvature of the total trapping potential near $\mathbf{T}_2$
\begin{equation}
\begin{aligned}
H_T = &\frac{m}{Q}\,\mathcal{R}(\phi_x, \phi_y, \phi_z)\,\left(\boldsymbol{\omega}\,\mathbbm{1}\,\boldsymbol{\omega}^T\right)\,\mathcal{R}(\phi_x, \phi_y, \phi_z)^T\\
= &
\begin{pmatrix}
280(17) & -16(22) & -53(6) \\
-16(22) & 133(7)& 19(20) \\
 -53(6) &  19(20) & 308(18) \\
\end{pmatrix}\,\mu\text{V}/(\mu\text{m})^2 \,\text{.}
\end{aligned}
\end{equation}
Here, $\mathbbm{1}$ denotes the three-dimensional identity matrix and errors correspond to systematic uncertainties of the angles.
The dominating systematic uncertainty in the mode configuration is caused by the uncertainty of the ion position $\Delta \mathbf{T}_2 = \pm$(1,1,5)\,$\mu$m and a related uncertainty in $\mathbf{E}_l$. 
Typical motional amplitudes after resonant excitation amount to $ |A_i| \approx 500\,$nm, corresponding to a coherent state with $\langle n_i \rangle \approx 1000$, while we calculate initial thermal states $\{\langle n_{1} \rangle, \langle n_{2} \rangle, \langle n_{3} \rangle\} \approx \{5, 6, 13\}$, assuming optimal Doppler cooling\,\cite{leibfried_quantum_2003}. 

\begin{figure}
\includegraphics[width=\columnwidth,keepaspectratio]{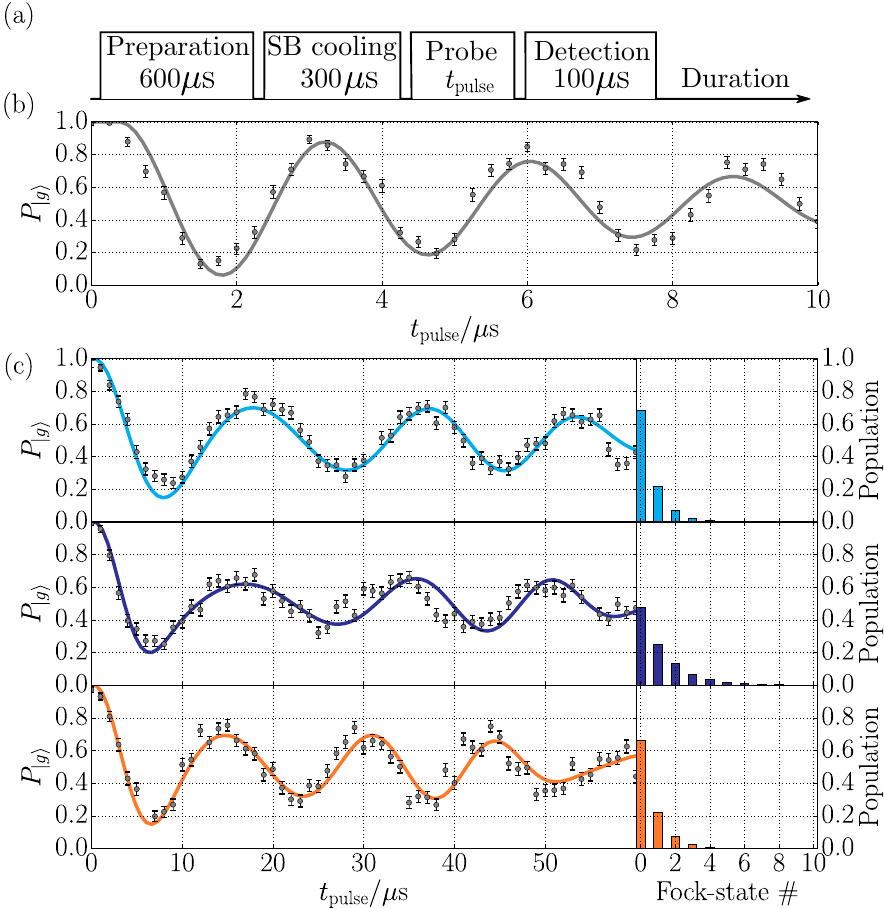}%
\caption{
(color online) Determination of mode orientations in the strong-binding limit. 
(a) After Doppler cooling and preparation of $|g\rangle$ and $\langle n_i \rangle \leq 1$ via resolved sideband cooling, we apply a coherent coupling of electronic and motional states with lasers BR and RR tuned to $\omega_{\rm{R}}$ for duration $t_{\rm{pulse}}$ and derive $P_{|g\rangle}$, the probability of detecting the ion in $|g\rangle$. 
Here,  $\boldsymbol{\omega}$ is determined by calibration measurements.
In subsequent measurements, we record $P_{|g\rangle}$ as a function of $t_{\rm{pulse}}$ for different couplings, where $\omega_{\rm{R}} \approx \omega_{\rm{s}}$ [see (b)] and $\omega_{\rm{R}} \approx \{\omega_{\rm{s}}+\omega_1, \omega_{\rm{s}}+\omega_2, \omega_{\rm{s}}+\omega_3\}$ [see (c)].
A combined model fit to the data (solid lines) and results from complementary measurements yield:  $\{ \phi_x, \phi_y,\phi_z\} = \{-9(2)_{\text{stat}}(5)_{\text{sys}},-51(2)_{\text{stat}}(5)_{\text{sys}},-15(1)_{\text{stat}}(5)_{\text{sys}}\}^{\circ}$, $\{\langle n_{1} \rangle, \langle n_{2} \rangle, \langle n_{3} \rangle\} = \{0.5(1)_{\rm{stat}}, 1.0(1)_{\rm{stat}}, 0.44(5)_{\rm{stat}}\}$.
Corresponding calculated thermal distributions $\mathcal{P}_i(n_i)$ of each mode are shown in bar charts on the right.
}
\label{fig4}
\end{figure}

Since mode configurations are subject to day-to-day variations of experimental parameters, e.g., due to stray potentials, we exploit our first method to perform a reference measurement and obtain $\{\phi_x, \phi_y,\phi_z\}= \{1(3)_{\text{stat}}(7)_{\text{sys}}, -52(3)_{\text{stat}}(6)_{\text{sys}}, -12(3)_{\text{stat}}(7)_{\text{sys}}\}^{\circ}$. 
We apply our second method, using the same single ion near $\mathbf{T}_2$ to avoid changes caused by intermittent loading, to determine the mode orientation by the following experimental sequence [see Fig.\,\ref{fig4}(a)]: 
After preparation of $|g\rangle$, we use resolved sideband cooling\,\cite{diedrich_laser_1989} to prepare all motional states close to their ground states. 
Then we couple the electronic and motional states with BR/RR along $\mathbf{\Delta k}$ for variable pulse duration $t_{\rm{pulse}}$, and derive the final population $P_{|g\rangle}$.
We set $\boldsymbol{\omega}/(2\pi) = \{ 3.76, 4.54, 5.76\}\,$MHz, determined by calibration measurements, and in subsequent experiments we probe different couplings at $\omega_{\rm{R}} \approx \{\omega_{\rm{s}}, \omega_{\rm{s}}+\omega_1, \omega_{\rm{s}}+\omega_2, \omega_{\rm{s}}+\omega_3\}$.
We show corresponding results of $P_{|g\rangle}$ as a function of $t_{\rm{pulse}}$ in Fig.\,\ref{fig4}(b).
%
%
For the analysis of our data, we use
\begin{equation}
\begin{aligned}
P_{\left|g \right\rangle}(t) =&\displaystyle\sum_{n_i = 0}^{n_{i,\rm{max}}}\left(\displaystyle\prod_{i=1}^{3}\mathcal{P}_i(n_i)\right)\cos^2\left(\frac{\Omega_{\rm{s}}}{2}t\right),
\label{propdown}
\end{aligned}
\end{equation}
%
%
where the motional-sensitive Rabi rate is given by:
\begin{equation}
\begin{aligned}
\Omega_{\rm{s}}=&\,\left\langle e_{\text{s}},n_1',n_2',n_3'\right|e^{i\langle\mathbf{\Delta k},\mathbf{u}_i\rangle}\left|g,n_1 ,n_2,n_3\right\rangle \\
\approx&\,\Omega_0 \cdot \displaystyle\prod_{i=1}^{3}\exp\left(\frac{-\eta_{i}^2}{2}\right)\eta_i^{|\Delta n_i|}\sqrt{\frac{n_{i<}!}{n_{i>}!}} L_{n_{i<}}^{|\Delta n_i|}(\eta_{i}^2).
\end{aligned}
\end{equation}
%
%
Here, $n_{i<}$ ($n_{i>}$) is the lesser (greater) of $n_i' = n_i + \Delta n_i$ and $n_i$, while $\Delta n_i$ denotes the change in the phonon number, i.e. the order of the sideband transition.
Further, $L_{n_{i<}}^{|\Delta n_i|}$ is the generalized Laguerre polynomial and $\Omega_0$ the motional-independent Rabi rate\,\cite{leibfried_quantum_2003}, tuned by beam parameters of BR and RR. 
Note, since we currently probe along $\mathbf{\Delta k}$ only, mode orientations are not uniquely determined and two out of three angles are derived.
Therefore, we iteratively fix one of the angles in combined fits to all datasets, using our reference measurements and average the outcomes.
For the analysis, we add a term in Eq.\,\eqref{propdown} to account for a decoherence rate $\Gamma_{\rm{dec}}$, truncate the sum in Eq.\,\eqref{propdown} at  $n_{i,\text{max}} = 11$, and assume thermal distributions of $\mathcal{P}_i(n_i)$. 
We find $\{ \phi_x, \phi_y,\phi_z\} = \{-9(2)_{\text{stat}}(5)_{\text{sys}},-51(2)_{\text{stat}}(5)_{\text{sys}},-15(1)_{\text{stat}}(5)_{\text{sys}}\}^{\circ}$, $\{\langle n_{1} \rangle, \langle n_{2} \rangle, \langle n_{3} \rangle\} = \{0.5(1)_{\rm{stat}}, 1.0(1)_{\rm{stat}}, 0.44(5)_{\rm{stat}}\}$, $\Omega_0/(2\pi) = 390(3)_{\rm{stat}}\,$kHz, and $\Gamma_{\text{dec}}/(2\pi) = 13(3)_{\rm{stat}}$\,kHz. 
Here, the systematic uncertainty is dominated by the limited knowledge of the orientation of $\Delta\mathbf{k}$ by $\pm5^{\circ}$.

%
To summarize, we introduce two general methods to fully characterize mode configurations of single ions.
To allow for detection of arbitrary mode configurations, each method needs to be applied from multiple different directions or can be supplemented by each other.
Further, our methods can be readily extended to characterization of mode configurations of multiple ions.
They may serve as standard procedures in experiments, where precise control of motional modes is inevitable, e.g., in two-dimensional ion trap arrays, used for quantum simulations\,\cite{schneider_experimental_2012,wilson_tunable_2014,mielenz_freely_2015}.
Here, tuning the motional parameters in real time permits state preparation of sympathetically cooled ions at all sites simultaneously as well as setting inter-site Coulomb interactions.
The latter has been demonstrated to mediate an effective interaction, that depends on the electronic states of the constituents and can be used either as spin-spin interaction or as data bus between qubits\,\cite{leibfried_experimental_2003,porras_quantum_2006,friedenauer_simulating_2008,harlander_trapped-ion_2011,wilson_tunable_2014}.
In addition, it may directly enable simulation of bosonic particles via phonon tunneling between neighboring sites defined by the ions, while corresponding local mode configurations define tunneling rates, pathways and related phases\,\cite{bermudez_synthetic_2011,*bermudez_photon-assisted-tunneling_2012}.
Moreover, adjusting mode configurations can aid either exploiting a more robust interaction bus\,\cite{bermudez_dissipation-assisted_2013} or investigating the sources of anomalous heating\,\cite{deslauriers_scaling_2006,*allcock_heating_2011,*daniilidis_surface_2014,*brownnutt_ion-trap_2015}, since recent experiments found a dependency of heating rates on the orientation of motional-mode vectors relative to the electrode structures\,\cite{schindler_polarization_2015}.
Further, our methods may be deployed to benchmark application of appropriate control potentials up to second order and, therefore, to compensate stray curvatures\,\cite{mielenz_freely_2015}.
In particular, in experiments with optical ion traps\,\cite{schneider_optical_2010,*huber_far-off-resonance_2014} or Rydberg states of ions \cite{muller_trapped_2008,*feldker_rydberg_2015}, precise adjustment of curvatures is necessary, maximizing trapping efficiencies and durations.
%
\vspace{3mm}

This work was supported by DFG (SCHA 972/6-1) and the Freiburg Institute for Advanced Studies (FRIAS). We thank J. Denter for technical assistance, Y.~Minet and L.~Nitzsche for helpful comments to the manuscript. 

\begin{thebibliography}{50}%
\makeatletter
\providecommand \@ifxundefined [1]{%
 \@ifx{#1\undefined}
}%
\providecommand \@ifnum [1]{%
 \ifnum #1\expandafter \@firstoftwo
 \else \expandafter \@secondoftwo
 \fi
}%
\providecommand \@ifx [1]{%
 \ifx #1\expandafter \@firstoftwo
 \else \expandafter \@secondoftwo
 \fi
}%
\providecommand \natexlab [1]{#1}%
\providecommand \enquote  [1]{``#1''}%
\providecommand \bibnamefont  [1]{#1}%
\providecommand \bibfnamefont [1]{#1}%
\providecommand \citenamefont [1]{#1}%
\providecommand \href@noop [0]{\@secondoftwo}%
\providecommand \href [0]{\begingroup \@sanitize@url \@href}%
\providecommand \@href[1]{\@@startlink{#1}\@@href}%
\providecommand \@@href[1]{\endgroup#1\@@endlink}%
\providecommand \@sanitize@url [0]{\catcode `\\12\catcode `\$12\catcode
  `\&12\catcode `\#12\catcode `\^12\catcode `\_12\catcode `\%12\relax}%
\providecommand \@@startlink[1]{}%
\providecommand \@@endlink[0]{}%
\providecommand \url  [0]{\begingroup\@sanitize@url \@url }%
\providecommand \@url [1]{\endgroup\@href {#1}{\urlprefix }}%
\providecommand \urlprefix  [0]{URL }%
\providecommand \Eprint [0]{\href }%
\providecommand \doibase [0]{http://dx.doi.org/}%
\providecommand \selectlanguage [0]{\@gobble}%
\providecommand \bibinfo  [0]{\@secondoftwo}%
\providecommand \bibfield  [0]{\@secondoftwo}%
\providecommand \translation [1]{[#1]}%
\providecommand \BibitemOpen [0]{}%
\providecommand \bibitemStop [0]{}%
\providecommand \bibitemNoStop [0]{.\EOS\space}%
\providecommand \EOS [0]{\spacefactor3000\relax}%
\providecommand \BibitemShut  [1]{\csname bibitem#1\endcsname}%
\let\auto@bib@innerbib\@empty
\bibitem [{\citenamefont {Feynman}(1982)}]{feynman_simulating_1982}%
  \BibitemOpen
  \bibfield  {author} {\bibinfo {author} {\bibfnamefont {R.~P.}\ \bibnamefont
  {Feynman}},\ }\href {\doibase 10.1007/BF02650179} {\bibfield  {journal}
  {\bibinfo  {journal} {Int. J. Theor. Phys.}\ }\textbf {\bibinfo {volume}
  {21}},\ \bibinfo {pages} {467} (\bibinfo {year} {1982})}\BibitemShut
  {NoStop}%
\bibitem [{\citenamefont {Buluta}\ and\ \citenamefont
  {Nori}(2009)}]{buluta_quantum_2009}%
  \BibitemOpen
  \bibfield  {author} {\bibinfo {author} {\bibfnamefont {I.}~\bibnamefont
  {Buluta}}\ and\ \bibinfo {author} {\bibfnamefont {F.}~\bibnamefont {Nori}},\
  }\href {\doibase 10.1126/science.1177838} {\bibfield  {journal} {\bibinfo
  {journal} {Science}\ }\textbf {\bibinfo {volume} {326}},\ \bibinfo {pages}
  {108} (\bibinfo {year} {2009})}\BibitemShut {NoStop}%
\bibitem [{\citenamefont {Schaetz}\ \emph {et~al.}(2013)\citenamefont
  {Schaetz}, \citenamefont {Monroe},\ and\ \citenamefont
  {Esslinger}}]{schaetz_focus_2013}%
  \BibitemOpen
  \bibfield  {author} {\bibinfo {author} {\bibfnamefont {T.}~\bibnamefont
  {Schaetz}}, \bibinfo {author} {\bibfnamefont {C.~R.}\ \bibnamefont {Monroe}},
  \ and\ \bibinfo {author} {\bibfnamefont {T.}~\bibnamefont {Esslinger}},\
  }\href {\doibase 10.1088/1367-2630/15/8/085009} {\bibfield  {journal}
  {\bibinfo  {journal} {New J. Phys.}\ }\textbf {\bibinfo {volume} {15}},\
  \bibinfo {pages} {085009} (\bibinfo {year} {2013})}\BibitemShut {NoStop}%
\bibitem [{\citenamefont {Georgescu}\ \emph {et~al.}(2014)\citenamefont
  {Georgescu}, \citenamefont {Ashhab},\ and\ \citenamefont
  {Nori}}]{georgescu_quantum_2014}%
  \BibitemOpen
  \bibfield  {author} {\bibinfo {author} {\bibfnamefont {I.}~\bibnamefont
  {Georgescu}}, \bibinfo {author} {\bibfnamefont {S.}~\bibnamefont {Ashhab}}, \
  and\ \bibinfo {author} {\bibfnamefont {F.}~\bibnamefont {Nori}},\ }\href
  {\doibase 10.1103/RevModPhys.86.153} {\bibfield  {journal} {\bibinfo
  {journal} {Rev. Mod. Phys.}\ }\textbf {\bibinfo {volume} {86}},\ \bibinfo
  {pages} {153} (\bibinfo {year} {2014})}\BibitemShut {NoStop}%
\bibitem [{\citenamefont {Cirac}\ and\ \citenamefont
  {Zoller}(2012)}]{cirac_goals_2012}%
  \BibitemOpen
  \bibfield  {author} {\bibinfo {author} {\bibfnamefont {J.~I.}\ \bibnamefont
  {Cirac}}\ and\ \bibinfo {author} {\bibfnamefont {P.}~\bibnamefont {Zoller}},\
  }\href {\doibase 10.1038/nphys2275} {\bibfield  {journal} {\bibinfo
  {journal} {Nat. Phys.}\ }\textbf {\bibinfo {volume} {8}},\ \bibinfo {pages}
  {264} (\bibinfo {year} {2012})}\BibitemShut {NoStop}%
\bibitem [{\citenamefont {Xiang}\ \emph {et~al.}(2013)\citenamefont {Xiang},
  \citenamefont {Ashhab}, \citenamefont {You},\ and\ \citenamefont
  {Nori}}]{xiang_hybrid_2013}%
  \BibitemOpen
  \bibfield  {author} {\bibinfo {author} {\bibfnamefont {Z.-L.}\ \bibnamefont
  {Xiang}}, \bibinfo {author} {\bibfnamefont {S.}~\bibnamefont {Ashhab}},
  \bibinfo {author} {\bibfnamefont {J.~Q.}\ \bibnamefont {You}}, \ and\
  \bibinfo {author} {\bibfnamefont {F.}~\bibnamefont {Nori}},\ }\href {\doibase
  10.1103/RevModPhys.85.623} {\bibfield  {journal} {\bibinfo  {journal} {Rev.
  Mod. Phys.}\ }\textbf {\bibinfo {volume} {85}},\ \bibinfo {pages} {623}
  (\bibinfo {year} {2013})}\BibitemShut {NoStop}%
\bibitem [{\citenamefont {Eckert}\ \emph {et~al.}(2008)\citenamefont {Eckert},
  \citenamefont {Romero-Isart}, \citenamefont {Rodriguez}, \citenamefont
  {Lewenstein}, \citenamefont {Polzik},\ and\ \citenamefont
  {Sanpera}}]{eckert_quantum_2008}%
  \BibitemOpen
  \bibfield  {author} {\bibinfo {author} {\bibfnamefont {K.}~\bibnamefont
  {Eckert}}, \bibinfo {author} {\bibfnamefont {O.}~\bibnamefont
  {Romero-Isart}}, \bibinfo {author} {\bibfnamefont {M.}~\bibnamefont
  {Rodriguez}}, \bibinfo {author} {\bibfnamefont {M.}~\bibnamefont
  {Lewenstein}}, \bibinfo {author} {\bibfnamefont {E.~S.}\ \bibnamefont
  {Polzik}}, \ and\ \bibinfo {author} {\bibfnamefont {A.}~\bibnamefont
  {Sanpera}},\ }\href {\doibase 10.1038/nphys776} {\bibfield  {journal}
  {\bibinfo  {journal} {Nat. Phys.}\ }\textbf {\bibinfo {volume} {4}},\
  \bibinfo {pages} {50} (\bibinfo {year} {2008})}\BibitemShut {NoStop}%
\bibitem [{\citenamefont {Moessner}\ \emph {et~al.}(2000)\citenamefont
  {Moessner}, \citenamefont {Sondhi},\ and\ \citenamefont
  {Chandra}}]{moessner_two-dimensional_2000}%
  \BibitemOpen
  \bibfield  {author} {\bibinfo {author} {\bibfnamefont {R.}~\bibnamefont
  {Moessner}}, \bibinfo {author} {\bibfnamefont {S.~L.}\ \bibnamefont
  {Sondhi}}, \ and\ \bibinfo {author} {\bibfnamefont {P.}~\bibnamefont
  {Chandra}},\ }\href {\doibase 10.1103/PhysRevLett.84.4457} {\bibfield
  {journal} {\bibinfo  {journal} {Phys. Rev. Lett.}\ }\textbf {\bibinfo
  {volume} {84}},\ \bibinfo {pages} {4457} (\bibinfo {year}
  {2000})}\BibitemShut {NoStop}%
\bibitem [{\citenamefont {Lewenstein}\ \emph {et~al.}(2007)\citenamefont
  {Lewenstein}, \citenamefont {Sanpera}, \citenamefont {Ahufinger},
  \citenamefont {Damski}, \citenamefont {Sen~De},\ and\ \citenamefont
  {Sen}}]{lewenstein_ultracold_2007}%
  \BibitemOpen
  \bibfield  {author} {\bibinfo {author} {\bibfnamefont {M.}~\bibnamefont
  {Lewenstein}}, \bibinfo {author} {\bibfnamefont {A.}~\bibnamefont {Sanpera}},
  \bibinfo {author} {\bibfnamefont {V.}~\bibnamefont {Ahufinger}}, \bibinfo
  {author} {\bibfnamefont {B.}~\bibnamefont {Damski}}, \bibinfo {author}
  {\bibfnamefont {A.}~\bibnamefont {Sen~De}}, \ and\ \bibinfo {author}
  {\bibfnamefont {U.}~\bibnamefont {Sen}},\ }\href {\doibase
  10.1080/00018730701223200} {\bibfield  {journal} {\bibinfo  {journal} {Adv.
  Phys.}\ }\textbf {\bibinfo {volume} {56}},\ \bibinfo {pages} {243} (\bibinfo
  {year} {2007})}\BibitemShut {NoStop}%
\bibitem [{\citenamefont {Hartmann}\ \emph {et~al.}(2006)\citenamefont
  {Hartmann}, \citenamefont {Brandão},\ and\ \citenamefont
  {Plenio}}]{hartmann_strongly_2006}%
  \BibitemOpen
  \bibfield  {author} {\bibinfo {author} {\bibfnamefont {M.~J.}\ \bibnamefont
  {Hartmann}}, \bibinfo {author} {\bibfnamefont {F.~G. S.~L.}\ \bibnamefont
  {Brandão}}, \ and\ \bibinfo {author} {\bibfnamefont {M.~B.}\ \bibnamefont
  {Plenio}},\ }\href {\doibase 10.1038/nphys462} {\bibfield  {journal}
  {\bibinfo  {journal} {Nat. Phys.}\ }\textbf {\bibinfo {volume} {2}},\
  \bibinfo {pages} {849} (\bibinfo {year} {2006})}\BibitemShut {NoStop}%
\bibitem [{\citenamefont {Hanson}\ and\ \citenamefont
  {Awschalom}(2008)}]{hanson_coherent_2008}%
  \BibitemOpen
  \bibfield  {author} {\bibinfo {author} {\bibfnamefont {R.}~\bibnamefont
  {Hanson}}\ and\ \bibinfo {author} {\bibfnamefont {D.~D.}\ \bibnamefont
  {Awschalom}},\ }\href {\doibase 10.1038/nature07129} {\bibfield  {journal}
  {\bibinfo  {journal} {Nat.}\ }\textbf {\bibinfo {volume} {453}},\ \bibinfo
  {pages} {1043} (\bibinfo {year} {2008})}\BibitemShut {NoStop}%
\bibitem [{\citenamefont {Porras}\ and\ \citenamefont
  {Cirac}(2006)}]{porras_quantum_2006}%
  \BibitemOpen
  \bibfield  {author} {\bibinfo {author} {\bibfnamefont {D.}~\bibnamefont
  {Porras}}\ and\ \bibinfo {author} {\bibfnamefont {J.~I.}\ \bibnamefont
  {Cirac}},\ }\href {\doibase 10.1103/PhysRevLett.96.250501} {\bibfield
  {journal} {\bibinfo  {journal} {Phys. Rev. Lett.}\ }\textbf {\bibinfo
  {volume} {96}},\ \bibinfo {pages} {250501} (\bibinfo {year}
  {2006})}\BibitemShut {NoStop}%
\bibitem [{\citenamefont {Schneider}\ \emph {et~al.}(2012)\citenamefont
  {Schneider}, \citenamefont {Porras},\ and\ \citenamefont
  {Schaetz}}]{schneider_experimental_2012}%
  \BibitemOpen
  \bibfield  {author} {\bibinfo {author} {\bibfnamefont {C.}~\bibnamefont
  {Schneider}}, \bibinfo {author} {\bibfnamefont {D.}~\bibnamefont {Porras}}, \
  and\ \bibinfo {author} {\bibfnamefont {T.}~\bibnamefont {Schaetz}},\ }\href
  {\doibase 10.1088/0034-4885/75/2/024401} {\bibfield  {journal} {\bibinfo
  {journal} {Rep. Prog. Phys.}\ }\textbf {\bibinfo {volume} {75}},\ \bibinfo
  {pages} {024401} (\bibinfo {year} {2012})}\BibitemShut {NoStop}%
\bibitem [{\citenamefont {Nowack}\ \emph {et~al.}(2007)\citenamefont {Nowack},
  \citenamefont {Koppens}, \citenamefont {Nazarov},\ and\ \citenamefont
  {Vandersypen}}]{nowack_coherent_2007}%
  \BibitemOpen
  \bibfield  {author} {\bibinfo {author} {\bibfnamefont {K.~C.}\ \bibnamefont
  {Nowack}}, \bibinfo {author} {\bibfnamefont {F.~H.~L.}\ \bibnamefont
  {Koppens}}, \bibinfo {author} {\bibfnamefont {Y.~V.}\ \bibnamefont
  {Nazarov}}, \ and\ \bibinfo {author} {\bibfnamefont {L.~M.~K.}\ \bibnamefont
  {Vandersypen}},\ }\href {\doibase 10.1126/science.1148092} {\bibfield
  {journal} {\bibinfo  {journal} {Science}\ }\textbf {\bibinfo {volume}
  {318}},\ \bibinfo {pages} {1430} (\bibinfo {year} {2007})}\BibitemShut
  {NoStop}%
\bibitem [{\citenamefont {Saffman}\ \emph {et~al.}(2010)\citenamefont
  {Saffman}, \citenamefont {Walker},\ and\ \citenamefont
  {M{\o}lmer}}]{saffman_quantum_2010}%
  \BibitemOpen
  \bibfield  {author} {\bibinfo {author} {\bibfnamefont {M.}~\bibnamefont
  {Saffman}}, \bibinfo {author} {\bibfnamefont {T.~G.}\ \bibnamefont {Walker}},
  \ and\ \bibinfo {author} {\bibfnamefont {K.}~\bibnamefont {M{\o}lmer}},\
  }\href {\doibase 10.1103/RevModPhys.82.2313} {\bibfield  {journal} {\bibinfo
  {journal} {Rev. Mod. Phys.}\ }\textbf {\bibinfo {volume} {82}},\ \bibinfo
  {pages} {2313} (\bibinfo {year} {2010})}\BibitemShut {NoStop}%
\bibitem [{\citenamefont {Deng}\ \emph {et~al.}(2010)\citenamefont {Deng},
  \citenamefont {Xie}, \citenamefont {Wu},\ and\ \citenamefont
  {Yang}}]{deng_storage_2010}%
  \BibitemOpen
  \bibfield  {author} {\bibinfo {author} {\bibfnamefont {Z.~J.}\ \bibnamefont
  {Deng}}, \bibinfo {author} {\bibfnamefont {Q.}~\bibnamefont {Xie}}, \bibinfo
  {author} {\bibfnamefont {C.~W.}\ \bibnamefont {Wu}}, \ and\ \bibinfo {author}
  {\bibfnamefont {W.~L.}\ \bibnamefont {Yang}},\ }\href {\doibase
  10.1103/PhysRevA.82.034306} {\bibfield  {journal} {\bibinfo  {journal} {Phys.
  Rev. A}\ }\textbf {\bibinfo {volume} {82}},\ \bibinfo {pages} {034306}
  (\bibinfo {year} {2010})}\BibitemShut {NoStop}%
\bibitem [{\citenamefont {Togan}\ \emph {et~al.}(2010)\citenamefont {Togan},
  \citenamefont {Chu}, \citenamefont {Trifonov}, \citenamefont {Jiang},
  \citenamefont {Maze}, \citenamefont {Childress}, \citenamefont {Dutt},
  \citenamefont {Sørensen}, \citenamefont {Hemmer}, \citenamefont {Zibrov},\
  and\ \citenamefont {Lukin}}]{togan_quantum_2010}%
  \BibitemOpen
  \bibfield  {author} {\bibinfo {author} {\bibfnamefont {E.}~\bibnamefont
  {Togan}}, \bibinfo {author} {\bibfnamefont {Y.}~\bibnamefont {Chu}}, \bibinfo
  {author} {\bibfnamefont {A.~S.}\ \bibnamefont {Trifonov}}, \bibinfo {author}
  {\bibfnamefont {L.}~\bibnamefont {Jiang}}, \bibinfo {author} {\bibfnamefont
  {J.}~\bibnamefont {Maze}}, \bibinfo {author} {\bibfnamefont {L.}~\bibnamefont
  {Childress}}, \bibinfo {author} {\bibfnamefont {M.~V.~G.}\ \bibnamefont
  {Dutt}}, \bibinfo {author} {\bibfnamefont {A.~S.}\ \bibnamefont {Sørensen}},
  \bibinfo {author} {\bibfnamefont {P.~R.}\ \bibnamefont {Hemmer}}, \bibinfo
  {author} {\bibfnamefont {A.~S.}\ \bibnamefont {Zibrov}}, \ and\ \bibinfo
  {author} {\bibfnamefont {M.~D.}\ \bibnamefont {Lukin}},\ }\href {\doibase
  10.1038/nature09256} {\bibfield  {journal} {\bibinfo  {journal} {Nat.}\
  }\textbf {\bibinfo {volume} {466}},\ \bibinfo {pages} {730} (\bibinfo {year}
  {2010})}\BibitemShut {NoStop}%
\bibitem [{\citenamefont {Daniilidis}\ \emph {et~al.}(2011)\citenamefont
  {Daniilidis}, \citenamefont {Narayanan}, \citenamefont {M\"oller},
  \citenamefont {Clark}, \citenamefont {Lee}, \citenamefont {Leek},
  \citenamefont {Wallraff}, \citenamefont {Schulz}, \citenamefont
  {Schmidt-Kaler},\ and\ \citenamefont
  {H\"affner}}]{daniilidis_fabrication_2011}%
  \BibitemOpen
  \bibfield  {author} {\bibinfo {author} {\bibfnamefont {N.}~\bibnamefont
  {Daniilidis}}, \bibinfo {author} {\bibfnamefont {S.}~\bibnamefont
  {Narayanan}}, \bibinfo {author} {\bibfnamefont {S.~A.}\ \bibnamefont
  {M\"oller}}, \bibinfo {author} {\bibfnamefont {R.}~\bibnamefont {Clark}},
  \bibinfo {author} {\bibfnamefont {T.~E.}\ \bibnamefont {Lee}}, \bibinfo
  {author} {\bibfnamefont {P.~J.}\ \bibnamefont {Leek}}, \bibinfo {author}
  {\bibfnamefont {A.}~\bibnamefont {Wallraff}}, \bibinfo {author}
  {\bibfnamefont {S.}~\bibnamefont {Schulz}}, \bibinfo {author} {\bibfnamefont
  {F.}~\bibnamefont {Schmidt-Kaler}}, \ and\ \bibinfo {author} {\bibfnamefont
  {H.}~\bibnamefont {H\"affner}},\ }\href {\doibase
  10.1088/1367-2630/13/1/013032} {\bibfield  {journal} {\bibinfo  {journal}
  {New J. Phys.}\ }\textbf {\bibinfo {volume} {13}},\ \bibinfo {pages} {013032}
  (\bibinfo {year} {2011})}\BibitemShut {NoStop}%
\bibitem [{\citenamefont {Sterling}\ \emph {et~al.}(2014)\citenamefont
  {Sterling}, \citenamefont {Rattanasonti}, \citenamefont {Weidt},
  \citenamefont {Lake}, \citenamefont {Srinivasan}, \citenamefont {Webster},
  \citenamefont {Kraft},\ and\ \citenamefont
  {Hensinger}}]{sterling_fabrication_2014}%
  \BibitemOpen
  \bibfield  {author} {\bibinfo {author} {\bibfnamefont {R.~C.}\ \bibnamefont
  {Sterling}}, \bibinfo {author} {\bibfnamefont {H.}~\bibnamefont
  {Rattanasonti}}, \bibinfo {author} {\bibfnamefont {S.}~\bibnamefont {Weidt}},
  \bibinfo {author} {\bibfnamefont {K.}~\bibnamefont {Lake}}, \bibinfo {author}
  {\bibfnamefont {P.}~\bibnamefont {Srinivasan}}, \bibinfo {author}
  {\bibfnamefont {S.~C.}\ \bibnamefont {Webster}}, \bibinfo {author}
  {\bibfnamefont {M.}~\bibnamefont {Kraft}}, \ and\ \bibinfo {author}
  {\bibfnamefont {W.~K.}\ \bibnamefont {Hensinger}},\ }\href {\doibase
  10.1038/ncomms4637} {\bibfield  {journal} {\bibinfo  {journal} {Nat. Comm.}\
  }\textbf {\bibinfo {volume} {5}},\ \bibinfo {pages} {3637} (\bibinfo {year}
  {2014})}\BibitemShut {NoStop}%
\bibitem [{\citenamefont {Seidelin}\ \emph {et~al.}(2006)\citenamefont
  {Seidelin}, \citenamefont {Chiaverini}, \citenamefont {Reichle},
  \citenamefont {Bollinger}, \citenamefont {Leibfried}, \citenamefont
  {Britton}, \citenamefont {Wesenberg}, \citenamefont {Blakestad},
  \citenamefont {Epstein}, \citenamefont {Hume}, \citenamefont {Itano},
  \citenamefont {Jost}, \citenamefont {Langer}, \citenamefont {Ozeri},
  \citenamefont {Shiga},\ and\ \citenamefont
  {Wineland}}]{seidelin_microfabricated_2006}%
  \BibitemOpen
  \bibfield  {author} {\bibinfo {author} {\bibfnamefont {S.}~\bibnamefont
  {Seidelin}}, \bibinfo {author} {\bibfnamefont {J.}~\bibnamefont
  {Chiaverini}}, \bibinfo {author} {\bibfnamefont {R.}~\bibnamefont {Reichle}},
  \bibinfo {author} {\bibfnamefont {J.~J.}\ \bibnamefont {Bollinger}}, \bibinfo
  {author} {\bibfnamefont {D.}~\bibnamefont {Leibfried}}, \bibinfo {author}
  {\bibfnamefont {J.}~\bibnamefont {Britton}}, \bibinfo {author} {\bibfnamefont
  {J.~H.}\ \bibnamefont {Wesenberg}}, \bibinfo {author} {\bibfnamefont {R.~B.}\
  \bibnamefont {Blakestad}}, \bibinfo {author} {\bibfnamefont {R.~J.}\
  \bibnamefont {Epstein}}, \bibinfo {author} {\bibfnamefont {D.~B.}\
  \bibnamefont {Hume}}, \bibinfo {author} {\bibfnamefont {W.~M.}\ \bibnamefont
  {Itano}}, \bibinfo {author} {\bibfnamefont {J.~D.}\ \bibnamefont {Jost}},
  \bibinfo {author} {\bibfnamefont {C.}~\bibnamefont {Langer}}, \bibinfo
  {author} {\bibfnamefont {R.}~\bibnamefont {Ozeri}}, \bibinfo {author}
  {\bibfnamefont {N.}~\bibnamefont {Shiga}}, \ and\ \bibinfo {author}
  {\bibfnamefont {D.~J.}\ \bibnamefont {Wineland}},\ }\href {\doibase
  10.1103/PhysRevLett.96.253003} {\bibfield  {journal} {\bibinfo  {journal}
  {Phys. Rev. Lett.}\ }\textbf {\bibinfo {volume} {96}},\ \bibinfo {pages}
  {253003} (\bibinfo {year} {2006})}\BibitemShut {NoStop}%
\bibitem [{\citenamefont {Schmied}\ \emph {et~al.}(2009)\citenamefont
  {Schmied}, \citenamefont {Wesenberg},\ and\ \citenamefont
  {Leibfried}}]{schmied_optimal_2009}%
  \BibitemOpen
  \bibfield  {author} {\bibinfo {author} {\bibfnamefont {R.}~\bibnamefont
  {Schmied}}, \bibinfo {author} {\bibfnamefont {J.~H.}\ \bibnamefont
  {Wesenberg}}, \ and\ \bibinfo {author} {\bibfnamefont {D.}~\bibnamefont
  {Leibfried}},\ }\href {\doibase 10.1103/PhysRevLett.102.233002} {\bibfield
  {journal} {\bibinfo  {journal} {Phys. Rev. Lett.}\ }\textbf {\bibinfo
  {volume} {102}},\ \bibinfo {pages} {233002} (\bibinfo {year}
  {2009})}\BibitemShut {NoStop}%
\bibitem [{\citenamefont {Mielenz}\ \emph {et~al.}(2015)\citenamefont
  {Mielenz}, \citenamefont {Kalis}, \citenamefont {Wittemer}, \citenamefont
  {Hakelberg}, \citenamefont {Schmied}, \citenamefont {Blain}, \citenamefont
  {Maunz}, \citenamefont {Leibfried}, \citenamefont {Warring},\ and\
  \citenamefont {Schaetz}}]{mielenz_freely_2015}%
  \BibitemOpen
  \bibfield  {author} {\bibinfo {author} {\bibfnamefont {M.}~\bibnamefont
  {Mielenz}}, \bibinfo {author} {\bibfnamefont {H.}~\bibnamefont {Kalis}},
  \bibinfo {author} {\bibfnamefont {M.}~\bibnamefont {Wittemer}}, \bibinfo
  {author} {\bibfnamefont {F.}~\bibnamefont {Hakelberg}}, \bibinfo {author}
  {\bibfnamefont {R.}~\bibnamefont {Schmied}}, \bibinfo {author} {\bibfnamefont
  {M.}~\bibnamefont {Blain}}, \bibinfo {author} {\bibfnamefont
  {P.}~\bibnamefont {Maunz}}, \bibinfo {author} {\bibfnamefont
  {D.}~\bibnamefont {Leibfried}}, \bibinfo {author} {\bibfnamefont
  {U.}~\bibnamefont {Warring}}, \ and\ \bibinfo {author} {\bibfnamefont
  {T.}~\bibnamefont {Schaetz}},\ }\href {http://arxiv.org/abs/1512.03559}
  {\bibfield  {journal} {\bibinfo  {journal} {arXiv:1512.03559 [physics,
  physics:quant-ph]}\ } (\bibinfo {year} {2015})},\ \bibinfo {note} {arXiv:
  1512.03559}\BibitemShut {NoStop}%
\bibitem [{\citenamefont {Jefferts}\ \emph {et~al.}(1995)\citenamefont
  {Jefferts}, \citenamefont {Monroe}, \citenamefont {Bell},\ and\ \citenamefont
  {Wineland}}]{jefferts_coaxial-resonator-driven_1995}%
  \BibitemOpen
  \bibfield  {author} {\bibinfo {author} {\bibfnamefont {S.~R.}\ \bibnamefont
  {Jefferts}}, \bibinfo {author} {\bibfnamefont {C.}~\bibnamefont {Monroe}},
  \bibinfo {author} {\bibfnamefont {E.~W.}\ \bibnamefont {Bell}}, \ and\
  \bibinfo {author} {\bibfnamefont {D.~J.}\ \bibnamefont {Wineland}},\ }\href
  {\doibase 10.1103/PhysRevA.51.3112} {\bibfield  {journal} {\bibinfo
  {journal} {Phys. Rev. A}\ }\textbf {\bibinfo {volume} {51}},\ \bibinfo
  {pages} {3112} (\bibinfo {year} {1995})}\BibitemShut {NoStop}%
\bibitem [{\citenamefont {Gudjons}\ \emph {et~al.}(1997)\citenamefont
  {Gudjons}, \citenamefont {Seibert},\ and\ \citenamefont
  {Werth}}]{gudjons_influence_1997}%
  \BibitemOpen
  \bibfield  {author} {\bibinfo {author} {\bibfnamefont {T.}~\bibnamefont
  {Gudjons}}, \bibinfo {author} {\bibfnamefont {P.}~\bibnamefont {Seibert}}, \
  and\ \bibinfo {author} {\bibfnamefont {G.}~\bibnamefont {Werth}},\ }\href
  {\doibase 10.1007/s003400050250} {\bibfield  {journal} {\bibinfo  {journal}
  {App. Phys. B}\ }\textbf {\bibinfo {volume} {65}},\ \bibinfo {pages} {57}
  (\bibinfo {year} {1997})}\BibitemShut {NoStop}%
\bibitem [{\citenamefont {Leibfried}\ \emph
  {et~al.}(2003{\natexlab{a}})\citenamefont {Leibfried}, \citenamefont {Blatt},
  \citenamefont {Monroe},\ and\ \citenamefont
  {Wineland}}]{leibfried_quantum_2003}%
  \BibitemOpen
  \bibfield  {author} {\bibinfo {author} {\bibfnamefont {D.}~\bibnamefont
  {Leibfried}}, \bibinfo {author} {\bibfnamefont {R.}~\bibnamefont {Blatt}},
  \bibinfo {author} {\bibfnamefont {C.}~\bibnamefont {Monroe}}, \ and\ \bibinfo
  {author} {\bibfnamefont {D.}~\bibnamefont {Wineland}},\ }\href {\doibase
  10.1103/RevModPhys.75.281} {\bibfield  {journal} {\bibinfo  {journal} {Rev.
  Mod. Phys.}\ }\textbf {\bibinfo {volume} {75}},\ \bibinfo {pages} {281}
  (\bibinfo {year} {2003}{\natexlab{a}})}\BibitemShut {NoStop}%
\bibitem [{\citenamefont {Tabakov}\ \emph {et~al.}(2015)\citenamefont
  {Tabakov}, \citenamefont {Benito}, \citenamefont {Blain}, \citenamefont
  {Clark}, \citenamefont {Clark}, \citenamefont {Haltli}, \citenamefont
  {Maunz}, \citenamefont {Sterk}, \citenamefont {Tigges},\ and\ \citenamefont
  {Stick}}]{tabakov_assembling_2015}%
  \BibitemOpen
  \bibfield  {author} {\bibinfo {author} {\bibfnamefont {B.}~\bibnamefont
  {Tabakov}}, \bibinfo {author} {\bibfnamefont {F.}~\bibnamefont {Benito}},
  \bibinfo {author} {\bibfnamefont {M.}~\bibnamefont {Blain}}, \bibinfo
  {author} {\bibfnamefont {C.~R.}\ \bibnamefont {Clark}}, \bibinfo {author}
  {\bibfnamefont {S.}~\bibnamefont {Clark}}, \bibinfo {author} {\bibfnamefont
  {R.~A.}\ \bibnamefont {Haltli}}, \bibinfo {author} {\bibfnamefont
  {P.}~\bibnamefont {Maunz}}, \bibinfo {author} {\bibfnamefont {J.~D.}\
  \bibnamefont {Sterk}}, \bibinfo {author} {\bibfnamefont {C.}~\bibnamefont
  {Tigges}}, \ and\ \bibinfo {author} {\bibfnamefont {D.}~\bibnamefont
  {Stick}},\ }\href {\doibase 10.1103/PhysRevApplied.4.031001} {\bibfield
  {journal} {\bibinfo  {journal} {Phys.Rev. App.}\ }\textbf {\bibinfo {volume}
  {4}},\ \bibinfo {pages} {031001} (\bibinfo {year} {2015})}\BibitemShut
  {NoStop}%
\bibitem [{\citenamefont {Hucul}\ \emph {et~al.}(2008)\citenamefont {Hucul},
  \citenamefont {Yeo}, \citenamefont {Olmschenk}, \citenamefont {Monroe},
  \citenamefont {Hensinger},\ and\ \citenamefont
  {Rabchuk}}]{Hucul:2008:TAI:2016976.2016977}%
  \BibitemOpen
  \bibfield  {author} {\bibinfo {author} {\bibfnamefont {D.}~\bibnamefont
  {Hucul}}, \bibinfo {author} {\bibfnamefont {M.}~\bibnamefont {Yeo}}, \bibinfo
  {author} {\bibfnamefont {S.}~\bibnamefont {Olmschenk}}, \bibinfo {author}
  {\bibfnamefont {C.}~\bibnamefont {Monroe}}, \bibinfo {author} {\bibfnamefont
  {W.~K.}\ \bibnamefont {Hensinger}}, \ and\ \bibinfo {author} {\bibfnamefont
  {J.}~\bibnamefont {Rabchuk}},\ }\href
  {http://dl.acm.org/citation.cfm?id=2016976.2016977} {\bibfield  {journal}
  {\bibinfo  {journal} {Quant. Inf. Comput.}\ }\textbf {\bibinfo {volume}
  {8}},\ \bibinfo {pages} {501} (\bibinfo {year} {2008})}\BibitemShut {NoStop}%
\bibitem [{\citenamefont {Wesenberg}(2008)}]{wesenberg_electrostatics_2008}%
  \BibitemOpen
  \bibfield  {author} {\bibinfo {author} {\bibfnamefont {J.~H.}\ \bibnamefont
  {Wesenberg}},\ }\href {\doibase 10.1103/PhysRevA.78.063410} {\bibfield
  {journal} {\bibinfo  {journal} {Phys.Rev. A}\ }\textbf {\bibinfo {volume}
  {78}},\ \bibinfo {pages} {063410} (\bibinfo {year} {2008})}\BibitemShut
  {NoStop}%
\bibitem [{\citenamefont {Schmied}(2010)}]{schmied_electrostatics_2010}%
  \BibitemOpen
  \bibfield  {author} {\bibinfo {author} {\bibfnamefont {R.}~\bibnamefont
  {Schmied}},\ }\href {\doibase 10.1088/1367-2630/12/2/023038} {\bibfield
  {journal} {\bibinfo  {journal} {New J. Phys.}\ }\textbf {\bibinfo {volume}
  {12}},\ \bibinfo {pages} {023038} (\bibinfo {year} {2010})}\BibitemShut
  {NoStop}%
\bibitem [{\citenamefont {Friedenauer}\ \emph {et~al.}(2006)\citenamefont
  {Friedenauer}, \citenamefont {Markert}, \citenamefont {Schmitz},
  \citenamefont {Petersen}, \citenamefont {Kahra}, \citenamefont {Herrmann},
  \citenamefont {Udem}, \citenamefont {H\"ansch},\ and\ \citenamefont
  {Sch\"atz}}]{friedenauer_high_2006}%
  \BibitemOpen
  \bibfield  {author} {\bibinfo {author} {\bibfnamefont {A.}~\bibnamefont
  {Friedenauer}}, \bibinfo {author} {\bibfnamefont {F.}~\bibnamefont
  {Markert}}, \bibinfo {author} {\bibfnamefont {H.}~\bibnamefont {Schmitz}},
  \bibinfo {author} {\bibfnamefont {L.}~\bibnamefont {Petersen}}, \bibinfo
  {author} {\bibfnamefont {S.}~\bibnamefont {Kahra}}, \bibinfo {author}
  {\bibfnamefont {M.}~\bibnamefont {Herrmann}}, \bibinfo {author}
  {\bibfnamefont {T.}~\bibnamefont {Udem}}, \bibinfo {author} {\bibfnamefont
  {T.~W.}\ \bibnamefont {H\"ansch}}, \ and\ \bibinfo {author} {\bibfnamefont
  {T.}~\bibnamefont {Sch\"atz}},\ }\href {\doibase 10.1007/s00340-006-2274-2}
  {\bibfield  {journal} {\bibinfo  {journal} {App. Phys. B}\ }\textbf {\bibinfo
  {volume} {84}},\ \bibinfo {pages} {371} (\bibinfo {year} {2006})}\BibitemShut
  {NoStop}%
\bibitem [{kal()}]{kalis_motional_supplement_2016}%
  \BibitemOpen
  \href@noop {} {}\bibinfo {note} {See Supplemental Material for mathematical
  details.}\BibitemShut {Stop}%
\bibitem [{\citenamefont {DeVoe}\ \emph {et~al.}(1989)\citenamefont {DeVoe},
  \citenamefont {Hoffnagle},\ and\ \citenamefont {Brewer}}]{devoe_role_1989}%
  \BibitemOpen
  \bibfield  {author} {\bibinfo {author} {\bibfnamefont {R.~G.}\ \bibnamefont
  {DeVoe}}, \bibinfo {author} {\bibfnamefont {J.}~\bibnamefont {Hoffnagle}}, \
  and\ \bibinfo {author} {\bibfnamefont {R.~G.}\ \bibnamefont {Brewer}},\
  }\href {\doibase 10.1103/PhysRevA.39.4362} {\bibfield  {journal} {\bibinfo
  {journal} {Phys.Rev. A}\ }\textbf {\bibinfo {volume} {39}},\ \bibinfo {pages}
  {4362} (\bibinfo {year} {1989})}\BibitemShut {NoStop}%
\bibitem [{\citenamefont {Berkeland}\ \emph {et~al.}(1998)\citenamefont
  {Berkeland}, \citenamefont {Miller}, \citenamefont {Bergquist}, \citenamefont
  {Itano},\ and\ \citenamefont {Wineland}}]{berkeland_minimization_1998}%
  \BibitemOpen
  \bibfield  {author} {\bibinfo {author} {\bibfnamefont {D.~J.}\ \bibnamefont
  {Berkeland}}, \bibinfo {author} {\bibfnamefont {J.~D.}\ \bibnamefont
  {Miller}}, \bibinfo {author} {\bibfnamefont {J.~C.}\ \bibnamefont
  {Bergquist}}, \bibinfo {author} {\bibfnamefont {W.~M.}\ \bibnamefont
  {Itano}}, \ and\ \bibinfo {author} {\bibfnamefont {D.~J.}\ \bibnamefont
  {Wineland}},\ }\href {\doibase 10.1063/1.367318} {\bibfield  {journal}
  {\bibinfo  {journal} {J. App. Phys.}\ }\textbf {\bibinfo {volume} {83}},\
  \bibinfo {pages} {5025} (\bibinfo {year} {1998})}\BibitemShut {NoStop}%
\bibitem [{\citenamefont {Diedrich}\ \emph {et~al.}(1989)\citenamefont
  {Diedrich}, \citenamefont {Bergquist}, \citenamefont {Itano},\ and\
  \citenamefont {Wineland}}]{diedrich_laser_1989}%
  \BibitemOpen
  \bibfield  {author} {\bibinfo {author} {\bibfnamefont {F.}~\bibnamefont
  {Diedrich}}, \bibinfo {author} {\bibfnamefont {J.~C.}\ \bibnamefont
  {Bergquist}}, \bibinfo {author} {\bibfnamefont {W.~M.}\ \bibnamefont
  {Itano}}, \ and\ \bibinfo {author} {\bibfnamefont {D.~J.}\ \bibnamefont
  {Wineland}},\ }\href {\doibase 10.1103/PhysRevLett.62.403} {\bibfield
  {journal} {\bibinfo  {journal} {Phys. Rev. Lett.}\ }\textbf {\bibinfo
  {volume} {62}},\ \bibinfo {pages} {403} (\bibinfo {year} {1989})}\BibitemShut
  {NoStop}%
\bibitem [{\citenamefont {Wilson}\ \emph {et~al.}(2014)\citenamefont {Wilson},
  \citenamefont {Colombe}, \citenamefont {Brown}, \citenamefont {Knill},
  \citenamefont {Leibfried},\ and\ \citenamefont
  {Wineland}}]{wilson_tunable_2014}%
  \BibitemOpen
  \bibfield  {author} {\bibinfo {author} {\bibfnamefont {A.~C.}\ \bibnamefont
  {Wilson}}, \bibinfo {author} {\bibfnamefont {Y.}~\bibnamefont {Colombe}},
  \bibinfo {author} {\bibfnamefont {K.~R.}\ \bibnamefont {Brown}}, \bibinfo
  {author} {\bibfnamefont {E.}~\bibnamefont {Knill}}, \bibinfo {author}
  {\bibfnamefont {D.}~\bibnamefont {Leibfried}}, \ and\ \bibinfo {author}
  {\bibfnamefont {D.~J.}\ \bibnamefont {Wineland}},\ }\href {\doibase
  10.1038/nature13565} {\bibfield  {journal} {\bibinfo  {journal} {Nat.}\
  }\textbf {\bibinfo {volume} {512}},\ \bibinfo {pages} {57} (\bibinfo {year}
  {2014})}\BibitemShut {NoStop}%
\bibitem [{\citenamefont {Leibfried}\ \emph
  {et~al.}(2003{\natexlab{b}})\citenamefont {Leibfried}, \citenamefont
  {DeMarco}, \citenamefont {Meyer}, \citenamefont {Lucas}, \citenamefont
  {Barrett}, \citenamefont {Britton}, \citenamefont {Itano}, \citenamefont
  {Jelenković}, \citenamefont {Langer}, \citenamefont {Rosenband},\ and\
  \citenamefont {Wineland}}]{leibfried_experimental_2003}%
  \BibitemOpen
  \bibfield  {author} {\bibinfo {author} {\bibfnamefont {D.}~\bibnamefont
  {Leibfried}}, \bibinfo {author} {\bibfnamefont {B.}~\bibnamefont {DeMarco}},
  \bibinfo {author} {\bibfnamefont {V.}~\bibnamefont {Meyer}}, \bibinfo
  {author} {\bibfnamefont {D.}~\bibnamefont {Lucas}}, \bibinfo {author}
  {\bibfnamefont {M.}~\bibnamefont {Barrett}}, \bibinfo {author} {\bibfnamefont
  {J.}~\bibnamefont {Britton}}, \bibinfo {author} {\bibfnamefont {W.~M.}\
  \bibnamefont {Itano}}, \bibinfo {author} {\bibfnamefont {B.}~\bibnamefont
  {Jelenković}}, \bibinfo {author} {\bibfnamefont {C.}~\bibnamefont {Langer}},
  \bibinfo {author} {\bibfnamefont {T.}~\bibnamefont {Rosenband}}, \ and\
  \bibinfo {author} {\bibfnamefont {D.~J.}\ \bibnamefont {Wineland}},\ }\href
  {\doibase 10.1038/nature01492} {\bibfield  {journal} {\bibinfo  {journal}
  {Nat.}\ }\textbf {\bibinfo {volume} {422}},\ \bibinfo {pages} {412} (\bibinfo
  {year} {2003}{\natexlab{b}})}\BibitemShut {NoStop}%
\bibitem [{\citenamefont {Friedenauer}\ \emph {et~al.}(2008)\citenamefont
  {Friedenauer}, \citenamefont {Schmitz}, \citenamefont {Glueckert},
  \citenamefont {Porras},\ and\ \citenamefont
  {Schaetz}}]{friedenauer_simulating_2008}%
  \BibitemOpen
  \bibfield  {author} {\bibinfo {author} {\bibfnamefont {A.}~\bibnamefont
  {Friedenauer}}, \bibinfo {author} {\bibfnamefont {H.}~\bibnamefont
  {Schmitz}}, \bibinfo {author} {\bibfnamefont {J.~T.}\ \bibnamefont
  {Glueckert}}, \bibinfo {author} {\bibfnamefont {D.}~\bibnamefont {Porras}}, \
  and\ \bibinfo {author} {\bibfnamefont {T.}~\bibnamefont {Schaetz}},\ }\href
  {\doibase 10.1038/nphys1032} {\bibfield  {journal} {\bibinfo  {journal} {Nat.
  Phys.}\ }\textbf {\bibinfo {volume} {4}},\ \bibinfo {pages} {757} (\bibinfo
  {year} {2008})}\BibitemShut {NoStop}%
\bibitem [{\citenamefont {Harlander}\ \emph {et~al.}(2011)\citenamefont
  {Harlander}, \citenamefont {Lechner}, \citenamefont {Brownnutt},
  \citenamefont {Blatt},\ and\ \citenamefont
  {H\"ansel}}]{harlander_trapped-ion_2011}%
  \BibitemOpen
  \bibfield  {author} {\bibinfo {author} {\bibfnamefont {M.}~\bibnamefont
  {Harlander}}, \bibinfo {author} {\bibfnamefont {R.}~\bibnamefont {Lechner}},
  \bibinfo {author} {\bibfnamefont {M.}~\bibnamefont {Brownnutt}}, \bibinfo
  {author} {\bibfnamefont {R.}~\bibnamefont {Blatt}}, \ and\ \bibinfo {author}
  {\bibfnamefont {W.}~\bibnamefont {H\"ansel}},\ }\href {\doibase
  10.1038/nature09800} {\bibfield  {journal} {\bibinfo  {journal} {Nat.}\
  }\textbf {\bibinfo {volume} {471}},\ \bibinfo {pages} {200} (\bibinfo {year}
  {2011})}\BibitemShut {NoStop}%
\bibitem [{\citenamefont {Bermudez}\ \emph {et~al.}(2011)\citenamefont
  {Bermudez}, \citenamefont {Schaetz},\ and\ \citenamefont
  {Porras}}]{bermudez_synthetic_2011}%
  \BibitemOpen
  \bibfield  {author} {\bibinfo {author} {\bibfnamefont {A.}~\bibnamefont
  {Bermudez}}, \bibinfo {author} {\bibfnamefont {T.}~\bibnamefont {Schaetz}}, \
  and\ \bibinfo {author} {\bibfnamefont {D.}~\bibnamefont {Porras}},\ }\href
  {\doibase 10.1103/PhysRevLett.107.150501} {\bibfield  {journal} {\bibinfo
  {journal} {Phys. Rev. Lett.}\ }\textbf {\bibinfo {volume} {107}},\ \bibinfo
  {pages} {150501} (\bibinfo {year} {2011})}\BibitemShut {NoStop}%
\bibitem [{\citenamefont {Bermudez}\ \emph {et~al.}(2012)\citenamefont
  {Bermudez}, \citenamefont {Schaetz},\ and\ \citenamefont
  {Porras}}]{bermudez_photon-assisted-tunneling_2012}%
  \BibitemOpen
  \bibfield  {author} {\bibinfo {author} {\bibfnamefont {A.}~\bibnamefont
  {Bermudez}}, \bibinfo {author} {\bibfnamefont {T.}~\bibnamefont {Schaetz}}, \
  and\ \bibinfo {author} {\bibfnamefont {D.}~\bibnamefont {Porras}},\ }\href
  {\doibase 10.1088/1367-2630/14/5/053049} {\bibfield  {journal} {\bibinfo
  {journal} {New J. Phys.}\ }\textbf {\bibinfo {volume} {14}},\ \bibinfo
  {pages} {053049} (\bibinfo {year} {2012})}\BibitemShut {NoStop}%
\bibitem [{\citenamefont {Bermudez}\ \emph {et~al.}(2013)\citenamefont
  {Bermudez}, \citenamefont {Schaetz},\ and\ \citenamefont
  {Plenio}}]{bermudez_dissipation-assisted_2013}%
  \BibitemOpen
  \bibfield  {author} {\bibinfo {author} {\bibfnamefont {A.}~\bibnamefont
  {Bermudez}}, \bibinfo {author} {\bibfnamefont {T.}~\bibnamefont {Schaetz}}, \
  and\ \bibinfo {author} {\bibfnamefont {M.~B.}\ \bibnamefont {Plenio}},\
  }\href {\doibase 10.1103/PhysRevLett.110.110502} {\bibfield  {journal}
  {\bibinfo  {journal} {Phys. Rev. Lett.}\ }\textbf {\bibinfo {volume} {110}},\
  \bibinfo {pages} {110502} (\bibinfo {year} {2013})}\BibitemShut {NoStop}%
\bibitem [{\citenamefont {Deslauriers}\ \emph {et~al.}(2006)\citenamefont
  {Deslauriers}, \citenamefont {Olmschenk}, \citenamefont {Stick},
  \citenamefont {Hensinger}, \citenamefont {Sterk},\ and\ \citenamefont
  {Monroe}}]{deslauriers_scaling_2006}%
  \BibitemOpen
  \bibfield  {author} {\bibinfo {author} {\bibfnamefont {L.}~\bibnamefont
  {Deslauriers}}, \bibinfo {author} {\bibfnamefont {S.}~\bibnamefont
  {Olmschenk}}, \bibinfo {author} {\bibfnamefont {D.}~\bibnamefont {Stick}},
  \bibinfo {author} {\bibfnamefont {W.~K.}\ \bibnamefont {Hensinger}}, \bibinfo
  {author} {\bibfnamefont {J.}~\bibnamefont {Sterk}}, \ and\ \bibinfo {author}
  {\bibfnamefont {C.}~\bibnamefont {Monroe}},\ }\href {\doibase
  10.1103/PhysRevLett.97.103007} {\bibfield  {journal} {\bibinfo  {journal}
  {Phys. Rev. Lett.}\ }\textbf {\bibinfo {volume} {97}},\ \bibinfo {pages}
  {103007} (\bibinfo {year} {2006})}\BibitemShut {NoStop}%
\bibitem [{\citenamefont {Allcock}\ \emph {et~al.}(2011)\citenamefont
  {Allcock}, \citenamefont {Harty}, \citenamefont {Janacek}, \citenamefont
  {Linke}, \citenamefont {Ballance}, \citenamefont {Steane}, \citenamefont
  {Lucas}, \citenamefont {Jarecki~Jr.}, \citenamefont {Habermehl},
  \citenamefont {Blain}, \citenamefont {Stick},\ and\ \citenamefont
  {Moehring}}]{allcock_heating_2011}%
  \BibitemOpen
  \bibfield  {author} {\bibinfo {author} {\bibfnamefont {D.~T.~C.}\
  \bibnamefont {Allcock}}, \bibinfo {author} {\bibfnamefont {T.~P.}\
  \bibnamefont {Harty}}, \bibinfo {author} {\bibfnamefont {H.~A.}\ \bibnamefont
  {Janacek}}, \bibinfo {author} {\bibfnamefont {N.~M.}\ \bibnamefont {Linke}},
  \bibinfo {author} {\bibfnamefont {C.~J.}\ \bibnamefont {Ballance}}, \bibinfo
  {author} {\bibfnamefont {A.~M.}\ \bibnamefont {Steane}}, \bibinfo {author}
  {\bibfnamefont {D.~M.}\ \bibnamefont {Lucas}}, \bibinfo {author}
  {\bibfnamefont {R.~L.}\ \bibnamefont {Jarecki~Jr.}}, \bibinfo {author}
  {\bibfnamefont {S.~D.}\ \bibnamefont {Habermehl}}, \bibinfo {author}
  {\bibfnamefont {M.~G.}\ \bibnamefont {Blain}}, \bibinfo {author}
  {\bibfnamefont {D.}~\bibnamefont {Stick}}, \ and\ \bibinfo {author}
  {\bibfnamefont {D.~L.}\ \bibnamefont {Moehring}},\ }\href {\doibase
  10.1007/s00340-011-4788-5} {\bibfield  {journal} {\bibinfo  {journal} {App.
  Phys. B}\ }\textbf {\bibinfo {volume} {107}},\ \bibinfo {pages} {913}
  (\bibinfo {year} {2011})}\BibitemShut {NoStop}%
\bibitem [{\citenamefont {Daniilidis}\ \emph {et~al.}(2014)\citenamefont
  {Daniilidis}, \citenamefont {Gerber}, \citenamefont {Bolloten}, \citenamefont
  {Ramm}, \citenamefont {Ransford}, \citenamefont {Ulin-Avila}, \citenamefont
  {Talukdar},\ and\ \citenamefont {H\"affner}}]{daniilidis_surface_2014}%
  \BibitemOpen
  \bibfield  {author} {\bibinfo {author} {\bibfnamefont {N.}~\bibnamefont
  {Daniilidis}}, \bibinfo {author} {\bibfnamefont {S.}~\bibnamefont {Gerber}},
  \bibinfo {author} {\bibfnamefont {G.}~\bibnamefont {Bolloten}}, \bibinfo
  {author} {\bibfnamefont {M.}~\bibnamefont {Ramm}}, \bibinfo {author}
  {\bibfnamefont {A.}~\bibnamefont {Ransford}}, \bibinfo {author}
  {\bibfnamefont {E.}~\bibnamefont {Ulin-Avila}}, \bibinfo {author}
  {\bibfnamefont {I.}~\bibnamefont {Talukdar}}, \ and\ \bibinfo {author}
  {\bibfnamefont {H.}~\bibnamefont {H\"affner}},\ }\href {\doibase
  10.1103/PhysRevB.89.245435} {\bibfield  {journal} {\bibinfo  {journal} {Phys.
  Rev. B}\ }\textbf {\bibinfo {volume} {89}},\ \bibinfo {pages} {245435}
  (\bibinfo {year} {2014})}\BibitemShut {NoStop}%
\bibitem [{\citenamefont {Brownnutt}\ \emph {et~al.}(2015)\citenamefont
  {Brownnutt}, \citenamefont {Kumph}, \citenamefont {Rabl},\ and\ \citenamefont
  {Blatt}}]{brownnutt_ion-trap_2015}%
  \BibitemOpen
  \bibfield  {author} {\bibinfo {author} {\bibfnamefont {M.}~\bibnamefont
  {Brownnutt}}, \bibinfo {author} {\bibfnamefont {M.}~\bibnamefont {Kumph}},
  \bibinfo {author} {\bibfnamefont {P.}~\bibnamefont {Rabl}}, \ and\ \bibinfo
  {author} {\bibfnamefont {R.}~\bibnamefont {Blatt}},\ }\href {\doibase
  10.1103/RevModPhys.87.1419} {\bibfield  {journal} {\bibinfo  {journal} {Rev.
  Mod. Phys.}\ }\textbf {\bibinfo {volume} {87}},\ \bibinfo {pages} {1419}
  (\bibinfo {year} {2015})}\BibitemShut {NoStop}%
\bibitem [{\citenamefont {Schindler}\ \emph {et~al.}(2015)\citenamefont
  {Schindler}, \citenamefont {Gorman}, \citenamefont {Daniilidis},\ and\
  \citenamefont {H\"affner}}]{schindler_polarization_2015}%
  \BibitemOpen
  \bibfield  {author} {\bibinfo {author} {\bibfnamefont {P.}~\bibnamefont
  {Schindler}}, \bibinfo {author} {\bibfnamefont {D.~J.}\ \bibnamefont
  {Gorman}}, \bibinfo {author} {\bibfnamefont {N.}~\bibnamefont {Daniilidis}},
  \ and\ \bibinfo {author} {\bibfnamefont {H.}~\bibnamefont {H\"affner}},\
  }\href {\doibase 10.1103/PhysRevA.92.013414} {\bibfield  {journal} {\bibinfo
  {journal} {Phys.Rev. A}\ }\textbf {\bibinfo {volume} {92}},\ \bibinfo {pages}
  {013414} (\bibinfo {year} {2015})}\BibitemShut {NoStop}%
\bibitem [{\citenamefont {Schneider}\ \emph {et~al.}(2010)\citenamefont
  {Schneider}, \citenamefont {Enderlein}, \citenamefont {Huber},\ and\
  \citenamefont {Schaetz}}]{schneider_optical_2010}%
  \BibitemOpen
  \bibfield  {author} {\bibinfo {author} {\bibfnamefont {C.}~\bibnamefont
  {Schneider}}, \bibinfo {author} {\bibfnamefont {M.}~\bibnamefont
  {Enderlein}}, \bibinfo {author} {\bibfnamefont {T.}~\bibnamefont {Huber}}, \
  and\ \bibinfo {author} {\bibfnamefont {T.}~\bibnamefont {Schaetz}},\ }\href
  {\doibase 10.1038/nphoton.2010.236} {\bibfield  {journal} {\bibinfo
  {journal} {Nat. Photonics}\ }\textbf {\bibinfo {volume} {4}},\ \bibinfo
  {pages} {772} (\bibinfo {year} {2010})}\BibitemShut {NoStop}%
\bibitem [{\citenamefont {Huber}\ \emph {et~al.}(2014)\citenamefont {Huber},
  \citenamefont {Lambrecht}, \citenamefont {Schmidt}, \citenamefont {Karpa},\
  and\ \citenamefont {Schaetz}}]{huber_far-off-resonance_2014}%
  \BibitemOpen
  \bibfield  {author} {\bibinfo {author} {\bibfnamefont {T.}~\bibnamefont
  {Huber}}, \bibinfo {author} {\bibfnamefont {A.}~\bibnamefont {Lambrecht}},
  \bibinfo {author} {\bibfnamefont {J.}~\bibnamefont {Schmidt}}, \bibinfo
  {author} {\bibfnamefont {L.}~\bibnamefont {Karpa}}, \ and\ \bibinfo {author}
  {\bibfnamefont {T.}~\bibnamefont {Schaetz}},\ }\href {\doibase
  10.1038/ncomms6587} {\bibfield  {journal} {\bibinfo  {journal} {Nat. Comm.}\
  }\textbf {\bibinfo {volume} {5}},\ \bibinfo {pages} {5587} (\bibinfo {year}
  {2014})}\BibitemShut {NoStop}%
\bibitem [{\citenamefont {M\"uller}\ \emph {et~al.}(2008)\citenamefont
  {M\"uller}, \citenamefont {Liang}, \citenamefont {Lesanovsky},\ and\
  \citenamefont {Zoller}}]{muller_trapped_2008}%
  \BibitemOpen
  \bibfield  {author} {\bibinfo {author} {\bibfnamefont {M.}~\bibnamefont
  {M\"uller}}, \bibinfo {author} {\bibfnamefont {L.}~\bibnamefont {Liang}},
  \bibinfo {author} {\bibfnamefont {I.}~\bibnamefont {Lesanovsky}}, \ and\
  \bibinfo {author} {\bibfnamefont {P.}~\bibnamefont {Zoller}},\ }\href
  {\doibase 10.1088/1367-2630/10/9/093009} {\bibfield  {journal} {\bibinfo
  {journal} {New J. Phys.}\ }\textbf {\bibinfo {volume} {10}},\ \bibinfo
  {pages} {093009} (\bibinfo {year} {2008})}\BibitemShut {NoStop}%
\bibitem [{\citenamefont {Feldker}\ \emph {et~al.}(2015)\citenamefont
  {Feldker}, \citenamefont {Bachor}, \citenamefont {Stappel}, \citenamefont
  {Kolbe}, \citenamefont {Gerritsma}, \citenamefont {Walz},\ and\ \citenamefont
  {Schmidt-Kaler}}]{feldker_rydberg_2015}%
  \BibitemOpen
  \bibfield  {author} {\bibinfo {author} {\bibfnamefont {T.}~\bibnamefont
  {Feldker}}, \bibinfo {author} {\bibfnamefont {P.}~\bibnamefont {Bachor}},
  \bibinfo {author} {\bibfnamefont {M.}~\bibnamefont {Stappel}}, \bibinfo
  {author} {\bibfnamefont {D.}~\bibnamefont {Kolbe}}, \bibinfo {author}
  {\bibfnamefont {R.}~\bibnamefont {Gerritsma}}, \bibinfo {author}
  {\bibfnamefont {J.}~\bibnamefont {Walz}}, \ and\ \bibinfo {author}
  {\bibfnamefont {F.}~\bibnamefont {Schmidt-Kaler}},\ }\href {\doibase
  10.1103/PhysRevLett.115.173001} {\bibfield  {journal} {\bibinfo  {journal}
  {Phys. Rev. Lett.}\ }\textbf {\bibinfo {volume} {115}},\ \bibinfo {pages}
  {173001} (\bibinfo {year} {2015})}\BibitemShut {NoStop}%
\end{thebibliography}
%

\end{document}